\documentclass[
  aps,
  prl,
  reprint,
  showpacs,
  groupedaddress,
  amsmath,
  amssymb,
  floatfix]{revtex4-1}
  
\usepackage{graphicx,epsfig,dcolumn,multirow}

\usepackage{color}

\newcolumntype{d}[1]{D{.}{.}{#1}}

\RequirePackage{xspace}

\newcommand{\dis}[1]{\begin{equation}\begin{split}#1\end{split}\end{equation}}
\newcommand{\be}{\begin{equation}}
\newcommand{\ee}{\end{equation}}
\def\bea{\begin{eqnarray}}
\def\eea{\end{eqnarray}}

\newcommand{\eq}[1]{Eq.~(\ref{#1})}

\newcommand{\VEV}[1]{\langle #1 \rangle}

\newcommand{\tev}{\,\textrm{TeV}}
\newcommand{\gev}{\,\textrm{GeV}}

\def\tb{\tan\beta}

\begin{document}

\title{\Large\bf 
Electroweak symmetry breaking by a neutral sector:
\\ 
Dynamical relaxation of the little hierarchy problem
}

\author{Bumseok Kyae\footnote{email: bkyae@pusan.ac.kr}
}
\affiliation{
Department of Physics, Pusan National University, Busan 609-735, Korea
}

\begin{abstract}

We propose a new dynamical relaxation mechanism  
of the little hierarchy problem,
based on a singlet extension of the 
minimal supersymmetric standard model (MSSM). 
In this scenario, the small soft mass 
parameter of an MSSM singlet 
is responsible for 
the electroweak symmetry breaking and 
the non-zero Higgs vacuum expectation value,  
whereas the effect of 
the large soft mass parameter of the Higgs boson, 
$-m_{h_u}^2$ is dynamically compensated 
by a flat direction of the MSSM singlets. 
The small singlet's soft mass and  
the Z boson mass can be protected, 
even if the stop mass is heavier than 10 or 20 TeV,
since the gravity-mediated supersymmetry breaking 
effects 
and 
the relevant Yukawa couplings are relatively small.   
%
%
A ``focus point'' of the singlet's soft mass parameter 
can emerge around the stop mass scale, 
and so various fine-tuning measures 
can reduce well below 100. 
Due to the relatively large gauge-mediated effects, 
the MSSM superpartners  
are much heavier than the experimental bounds, and 
the unwanted flavor changing processes are 
adequately suppressed.

\end{abstract}



\maketitle


One of the long standing problems 
in theoretical particle physics 
is the gauge hierarchy problem. 
It is basically a naturalness problem associated with 
the relatively small Higgs boson mass and 
the resulting electroweak (EW) interaction scale 
much lower than 
a ultraviolet (UV) cutoff energy scale, 
below which the standard model (SM) can be valid. 
%
%
%
For last four decades, the question 
how the small Higgs boson mass 
can naturally be maintained 
against the large quantum corrections
without a fine tuning
has encouraged many physicists to propose  
various UV theories embedding the SM 
just above the EW scale. 
%
Thus, a new physics has been expected to be present 
around the EW scale,   
by which the counter operators are provided   
to cancel the quadratic divergences
appearing in the radiative corrections 
to the Higgs mass parameter,  
and renormalize it.   
Otherwise, a fine-tuning associated with  
its renormalization becomes serious. 

In particular,  
introduction of supersymmetry (SUSY)
at the EW scale has been accepted as 
the most promising way to resolve 
the problem \cite{book,Witten}:  
In SUSY theories, 
the needed counter terms in the Lagrangian
are dynamically generated by superpartners 
at the SUSY scale.
In the minimal supersymmetric standard model (MSSM),
moreover, the renormalization group (RG) evolutions 
of the three SM gauge coupling constants 
turned out to be precisely unified 
around $10^{16} \gev$ energy scale,   
when the superpartners' contributions to them are 
included \cite{book,MSSMunif}.  
It might be an evidence of 
the presence of 
a theory unifying all the SM gauge interactions 
at that scale, 
and so the MSSM has been regarded as a guiding model   
leading to such a grand unified theory (GUT). 
%

In supergravity (SUGRA) models,  
the EW symmetry is radiatively broken at low energy 
through the RG effect on the Higgs soft mass 
parameter $m_{h_u}^2$ 
due to the large top quark Yukawa coupling. 
As seen in the following two extreme conditions 
of the scalar potential for the two Higgs bosons, 
$h_u$ and $h_d$ in the MSSM \cite{book,HiggsPot}, 
\bea \label{extrm1}
&&|\mu|^2+\frac12 M_Z^2=\frac{m_{h_d}^2-m_{h_u}^2{\rm tan}^2\beta}{{\rm tan}^2\beta-1} ,
\\ \label{extrm2}
&&2|B\mu|+ M_Z^2~{\rm sin}2\beta=(m_{h_u}^2-m_{h_d}^2)~{\rm tan}2\beta , 
\eea
non-zero Higgs vacuum expectation values (VEVs) 
and the Z boson mass
[$M_Z^2=(g_2^2+g_Y^2)(\VEV{h_u}^2+\VEV{h_d}^2)/2$]
are generated with $\pi/4<\beta<\pi/2$, 
when $m_{h_u}^2$ becomes negative 
via its RG evolution at low energies. 
Here $\mu$ and $B\mu$ denote the mass of 
the Higgsinos 
(superpartners of the Higgs scalars)  and 
its corresponding soft mass parameters (``B-term'') 
and ${\rm tan}\beta$ ($\equiv \VEV{h_u}/\VEV{h_d}$) is 
the ratio of the VEVs of 
the two Higgs doublets.  
In SUSY models, thus, $m_{h_{u,d}}^2$, 
$|\mu|^2$ and $M_Z^2$ are required to be of 
a similar size
for the naturalness of the Z boson mass.  
Although the energy scale the LHC probes 
has been raised higher and higher so far, however, 
any new physics signal 
has not be observed yet.    
It implies that 
as the UV cutoff scale of the SM 
becomes higher and higher,
the fine-tuning problem for 
the Higgs mass parameter 
is being serious more and more.
In fact, all the theoretical puzzles raised in the SM 
still remain unsolved at the moment.

A barometer of the naturalness of the MSSM is
the stop (superpartner of the top quark) mass:
A too heavy stop mass induces 
a large value of $m_{h_u}^2$, 
which requires a fine-tuning 
with other parameters in \eq{extrm1} 
to get the Z boson mass of $91 \gev$.  
However, the experimental stop (gluino) mass bound 
has been already 
exceeded $1 \tev$ ($2 \tev$) \cite{PDG}, 
by which a fine-tuning of sub-percent level
seems to be needed already. 
%
%
Moreover, the observed Higgs boson mass, 
$125 \gev$ \cite{PDG} 
is too heavy as a SUSY Higgs boson mass, 
because it requires a too heavy stop mass 
for explaining it. 
According to the recent theoretical analyses
based on three loop calculations, 
$10$--$20 \tev$ stop mass is necessary in the MSSM
for explaining the 125 GeV Higgs mass
without a quite large stop mixing effect \cite{3-loop}.  
Accordingly, a fine-tuning of order 
$10^{-3}$ or $10^{-4}$ seems to be unavoidable, 
even if there exists SUSY at $10$--$20 \tev$ scale.   
It is called ``little hierarchy problem.''   

%
%
Apart from such a fine-tuning problem, 
some phenomenological problems 
were also pointed out in two representative 
SUSY breaking scenarios \cite{book}.
In gravity mediation scenario, 
where all the scalar fields obtain  
SUSY breaking soft masses, 
sizable flavor changing neutral currents (FCNC)
are generically admitted.  
On the other hand, in gauge mediation, 
where only scalar fields carrying the SM gauge charges
acquire the soft masses at the leading order, 
it is hard to get the $\mu/B\mu$ terms
of desirable size, 
while the flavor problem doesn't arise. 
%
 
In this letter, we will discuss the possibility that 
the EW phase transition is triggered 
by a mass parameter of a singlet 
rather than $m_{h_u}^2$ in the SUSY framework. 
We will employ both the gravity- and gauge-mediated 
SUSY breaking scenarios, 
assuming the gauge-mediated effects 
dominate over the gravity-mediated ones  
such that 
all the super particles  
carrying the SM gauge quantum numbers 
are made much heavier than 
the experimental bounds 
and unwanted flavor changing processes are  
sufficiently suppressed.  
As mentioned above, 
$10$--$20 \tev$ stop mass could explain 
the observed Higgs mass well.  
On the other hand, the soft masses of SM singlets 
remain relatively small in this case, 
since they are generated 
only by the gravity mediation.  
Hence, the scale of the EW symmetry breaking 
can be much lower than the ordinary MSSM 
SUSY particles' mass scale.  
We could restore the traditional radiative EW symmetry
breaking scenario with it.



As a benchmark model, let us consider the following 
form of a singlet extension of the MSSM 
in the superpotential:
\dis{ \label{W}
W\supset \left(
\lambda_1 X+\lambda_2\phi+\mu\right)h_uh_d
+ MXY +\frac{\kappa}{2}Y\phi^2 .
}
where $X$, $Y$, and $\phi$ denote newly introduced
singlet superfields inert 
under the MSSM gauge interaction.
Here $M$ is a mass parameter of order $1$--$10 \tev$, 
while $\{\lambda_{1}, \lambda_2, \kappa\}$ are 
dimensionless coupling constants. 
As seen in the first two terms of \eq{W}, 
the MSSM $\mu$ term is promoted to 
the trilinear couplings among $X$, $\phi$, 
and the MSSM Higgs fields 
apart from the bare $\mu$ term.   
%
%
%
%
In \eq{W}, we ignored the existence of 
e.g. $\phi Y$ term, 
assuming its dimensionful SUSY coupling 
is small enough, 
because it is not crucial in our analysis.    
Nonetheless, 
this superpotential does not admit any 
accidental symmetry.
As will be seen below, instead,  
a flat direction is found in the SUSY limit 
in this model.  
\eq{W} could be a remnant of 
the U(1) Pecci-Quinn symmetry [$U(1)_{\rm PQ}$] 
breaking mechanism 
at an intermediate scale \cite{Kim-Nilles}.
 

%
%
\begin{table}[!h]
\begin{center}
\begin{tabular}
{c|cccccc} 
{\rm Superfields}  &   ~$X_{1,2}$~   &
 ~$Y$~  &  ~$Z$~ & ~$\phi$~   &  
 ~$\Psi$~ & ~$\Psi^c$ 
\\
\hline 
U(1)$_{\rm PQ}$  & ~$-1$~ & ~$4/3$~ & ~$5/6$~ 
& ~$-2/3$~ & ~$1/6$~ & ~$-1/6$~  
%
\end{tabular}
\end{center}\caption{Extra MSSM Singlet Superfields 
charged under U(1)$_{\rm PQ}$. 
The ordinary MSSM superfields   
including the two Higgs doublets 
should carry proper U(1)$_{\rm PQ}$ charges. 
}\label{tab:PQ}
\end{table}
As a UV model, one can consider, for instance,  
\bea \label{Wuv}
&&\qquad\qquad 
W_{\rm UV}\supset \Psi\left(y_1X_1+y_2X_2\right)Z
+y_3\Psi^cZ\phi
\\ \nonumber 
&&+\left(y_4X_1+y_5X_2\right)h_uh_d 
+\frac{(\Psi^{c})^2}{M_P}
\left(y_6X_1+y_7X_2\right)Y+\frac{\kappa}{2}Y\phi^2 ,
\eea
where $M_P$ denotes the reduced Planck mass 
($\approx 2.4\times 10^{18} \gev$), and 
two $X$s, i.e., $\{X_1, X_2\}$, and the spurion fields 
$\{\Psi, \Psi^c\}$ breaking the U(1)$_{\rm PQ}$ 
are introduced.
The global U(1)$_{\rm PQ}$ charge assignment is 
presented in TABLE \ref{tab:PQ}.  
We suppose that 
the scalar components of $\{\Psi, \Psi^c\}$ 
develop non-zero VEVs
at an intermediate scale 
inside the ``axion window'' \cite{axionRVW}, 
say, of order $10^{11} \gev$.
By non-zero 
VEVs of the scalar components of $\{\Psi, \Psi^c\}$, 
the U(1)$_{\rm PQ}$ is completely broken, 
and $Z$ and one combination of $X_{1,2}$
[$=(y_1X_1+y_2X_2)/\sqrt{y_1^2+y_2^2}\equiv X_H$] 
become superheavy.    
Integrating out such heavy superfields leaves 
\eq{W}, in which $X$ is identified with 
the light mode of $X_{1,2}$ orthogonal to $X_H$, 
and $\lambda_2$ is proportional to 
$y_3\langle\Psi^c\rangle/\langle\Psi\rangle$. 
The mass term of $X$ and $Y$ in \eq{W} is 
generated from the non-renormalizable term in \eq{Wuv} 
\cite{Kim-Nilles}. 
A mass term of $\phi$ and $Y$ is also induced. 
However, it turns out to leave intact the existence of  
the flat direction: It just deform it. 
We will ignore the term just for simplicity.  
In a similar way, the bare $\mu$ term in \eq{W} 
can also be generated with more spurion fields. 


The resulting scalar potential with \eq{W} 
is given by 
\bea \label{V}
&&V\supset
\Big|\lambda_1X+\lambda_2\phi+\mu\Big|^2|H|^2
~+~\left|\frac{\kappa}{2}\phi^2+MX\right|^2
\nonumber \\
&&\qquad 
+|\lambda_1 h_uh_d+MY|^2
+|\lambda_2h_uh_d+\kappa\phi Y|^2
\\
&&\qquad\quad ~~ 
+m_X^2|X|^2+m_Y^2|Y|^2+m_\phi^2|\phi|^2
\nonumber \\
&&+\left\{(
A_1X+A_2\phi
+B\mu)h_uh_d
+Mb XY+\frac{\kappa}{2}aY\phi^2
+{\rm h.c.}\right\} , 
\nonumber
\eea
where $|H|^2\equiv |h_u|^2+|h_d|^2$, and 
$\{m_{X,Y,\phi}^2, A_{1,2}, a, b, B\}$ are  
soft SUSY breaking parameters. 
Here we assume a hierarchy between $M$ and 
such soft parameters of the MSSM singlets, 
$|M|^2\gg |m_{X,Y,\phi}^2|, |A_{1,2}|^2, 
|a|^2, |b|^2, |B|^2$. 
In addition, we will regard these MSSM singlets' 
soft parameters as 
being relatively suppressed also than 
the ordinary MSSM soft (squared mass) parameters. 
It can be realized 
if the gravity-mediated SUSY breaking effects are 
relatively suppressed than the gauge-mediated ones.  
Here we note that 
a flat direction, $\kappa\phi^2/2+MX=0$ 
with $\langle h_u\rangle =\langle h_d\rangle 
=\langle Y\rangle=0$   
exists in the SUSY limit,  
since the Higgs gets a VEV only 
by a soft mass parameter 
as will be seen below.
Accordingly, the VEVs of $\phi$ and $X$ can be 
arbitrarily large in this limit. 
The flat direction is lifted 
only by small soft parameters.
From Eqs.~(\ref{W}) and (\ref{V}), 
the effective $\mu$ and $B\mu$ parameters 
read as follows: 
\bea \label{effmu}
&&\mu_{\rm eff}=\lambda_1\langle X\rangle
+\lambda_2\langle\phi\rangle +\mu ,
\\ \nonumber 
&&B\mu_{\rm eff}=\left(\lambda_1M^*
+\lambda_2\kappa^*\langle\phi\rangle^*\right)
\langle Y\rangle^* 
\\ \nonumber 
&&\qquad 
+A_1\langle X\rangle
+A_2\langle\phi\rangle +B\mu ,
\eea
which replace $\mu$ and $B\mu$ 
in Eqs.~(\ref{extrm1}) and (\ref{extrm2}) in our case. 
%
%



From \eq{V} the extreme conditions for $X$, $Y$, and $\phi$ are derived as follows:  
\bea \label{ExtrmCondi}
\left\{
\begin{array}{l}
{\cal M}_X^2X+M^*b^*Y^*
=-\frac{\kappa}{2}M^*\phi^2
-(\lambda_2\phi+\mu) \lambda_1^*|H|^2
\nonumber \\ 
\qquad\qquad\qquad\qquad\quad  
-A_1^*h_u^*h_d^* ~,
\nonumber \\
{\cal M}_Y^2Y^*+MbX
=-\frac{\kappa}{2}a\phi^2
-\left(\lambda_1^*M +\lambda_2^*\kappa\phi \right)h_u^*h_d^* ~,
\\
\big(|\kappa Y|^2+|\lambda_2H|^2+m_\phi^2\big)\phi
+\left(\frac{\kappa}{2}\phi^2+MX\right)\kappa^*\phi^*
\nonumber \\
\qquad\qquad\quad~ 
+(\lambda_1X+\mu) \lambda_2^*|H|^2
+A_2^*h_u^*h_d^*
\nonumber \\ 
\qquad\qquad\quad~ 
+(\lambda_2h_uh_d+a^*\phi^*)\kappa^*Y^* =0 .
\end{array}
\right.
\nonumber
\eea
For brevity, here, we introduced 
${\cal M}_X^2$ and ${\cal M}_Y^2$
defined as 
${\cal M}_X^2\equiv |\lambda_1H|^2+m_X^2+|M|^2$
($\approx |M|^2$) and ${\cal M}_Y^2\equiv 
|\kappa\phi|^2+m_Y^2+|M|^2$, respectively. 
Together with Eqs.~(\ref{extrm1}) and (\ref{extrm2}), 
thus, we should solve the five coupled equations 
in total. 
We should first note that
in a large limit of $\phi$ and $X$ 
($\gg |A_2|,~|\kappa Y|$)
with $\kappa\phi^2/2+MX\approx 0$,  
the VEV of Higgs, $H$ 
is constrained to roughly be of order 
$m_\phi/\lambda_{2}$, $\kappa Y$, or $a\kappa Y$  
from the third equation. 
%
%
We will see it more clearly below.


For a large enough $M$ and $\phi$, 
the solutions of $X$ and $Y$ 
to the first two equations in the above  
can approximately be expressed 
in terms of $\phi$ and $H$:
\bea \label{XY}
&&X\approx \frac{-\kappa\phi^2}{2{\cal M}_X^2}M^*
\left[1-\frac{(a-b)b^*}{{\cal M}_Y^2}
+\frac{2(\lambda_2\phi+\mu)\lambda_1^*|H|^2}
{\kappa\phi^2 M^*}\right] ,
\nonumber \\
&&Y^*\approx \frac{-\kappa\phi^2}{2{\cal M}_Y^2}
\left(a-b\right) 
-\frac{(\lambda_1^*M+\lambda_2^*\kappa\phi)
h_u^*h_d^*}{{\cal M}_Y^2} .
\eea
Then the flat direction, $(\kappa/2)\phi^2+MX=0$ 
is lifted to $(\kappa/2)\phi^2+MX\approx 
(\kappa\phi^2/2{\cal M}_X^2)\big[
|\lambda_1H|^2+m_X^2
+(|M|^2/{\cal M}_Y^2)(a-b)b^*
-(2M/\kappa\phi^2)
(\lambda_2\phi+\mu)\lambda_1^*|H|^2)\big]$. 
Plugging them into \eq{V}, 
the quartic terms of $\phi$ and  
the Higgs scalar such as 
$|(\kappa\phi^2/2{\cal M}_X^2)|\lambda_1H|^2|^2$
are induced. 
They are helpful 
for raising the Higgs boson mass. 

Inserting the above expressions into the third 
equation, 
we get the equation for $\phi$ or 
$T_\zeta$ ($\equiv \kappa\phi/M$):
\bea \label{phi}
&&\frac{|T_\zeta|^2}{2}
\left(|\lambda_1H|^2+m_X^2\right)
-T_\zeta^*\left[\lambda_2
+\frac{\mu}{\phi}\right]\lambda_1^*|H|^2 
-\frac{T_\zeta}{2}\lambda_2^*\lambda_1|H|^2 
\nonumber \\
&& ~ +\frac{\mu}{\phi}\lambda_2^*|H|^2
+\left(|\lambda_2H|^2+m_\phi^2\right)
\approx \frac{|T_\zeta|^2(|T_\zeta|^2+2)}{4(|T_\zeta|^2+1)^2}~|a-b|^2 , 
\nonumber \\ \label{phiH}
&& \quad ~{\rm or}~ \quad 
|H|^2\approx \frac{-m_\phi^2-\frac12\left(
m_X^2-|a-b|^2 f_T
\right)|T_\zeta|^2}
{\left(\lambda_2-\frac12\lambda_1T_\zeta
+\frac{\mu}{\phi}\right)
\left(\lambda_2^*-\lambda_1^*T_\zeta^*\right)} , 
\eea
unless $\langle\phi\rangle=0$. 
Here we set $f_T\equiv 
(1+|T_\zeta|^2/2)/(1+|T_\zeta|^2)^2$,  
which drops from $1$ to $0$  
as $|T_\zeta|$ increases.
We note here that $H$ can develop a nonzero VEV, 
when $m_\phi^2$ (and/or $m_X^2$) becomes negative 
for a positive denominator in \eq{phiH}.
Otherwise, $\langle \phi\rangle$ should be zero. 
Then, $\langle H\rangle$ should also vanish 
particularly for $|\mu|^2\gtrsim -m_{h_u}^2$. 
We will see later that 
$\langle\phi\rangle=\langle H\rangle=0$ 
is a saddle (stable) point 
when $m_\phi^2$ is negative (positive). 

We should note also that unlike in the MSSM
the size of $|\langle H\rangle|^2$ is basically 
of order $m_{\phi,X}^2/|\lambda_{2,1}|^2$ 
rather than ${\cal O}(m_{h_u}^2)$ 
for almost whole range of $T_\zeta$, 
only if $M$ and $\phi$ are large enough. 
In this model, therefore, their smallness is 
responsible for the smallness of the Higgs VEV and  
eventually the Z boson mass. 
As mentioned above, their smallness could be 
protected for relatively low scale of 
the gravity-mediated SUSY breaking, namely, 
$F_{\rm grav}/(\sqrt{3}M_P) \ll 
F_{g}/(16\pi^2\Lambda_{M})$. 
Here $F_{\rm grav}$ and $F_{g}$ denote 
the SUSY breaking sources in a hidden sector 
whose effects are mediated to the observable sector 
through the gravity and the SM gauge interactions, 
respectively, and 
$\Lambda_{M}$ stands for  
the messenger scale. 

Accordingly, 
the extreme condition of the Higgs fields, 
\eq{extrm1} should be met  
by the modulus-like field $\phi$:
$\langle \phi\rangle$ (and $\langle X\rangle$) 
should compensate the large value of 
$-m_{h_u}^2$ in \eq{extrm1}. 
As a result, 
the Higgsino mass $\mu_{\rm eff}$
is necessarily large in this model,
of order $|m_{h_u}^2|^{1/2}$.    
It is a salient feature of this model, 
distinguished from other SUSY models 
pursuing the naturalness \cite{FP,MyFP},   
or even the split SUSY model \cite{splitSUSY}. 
While $\mu_{\rm eff}$ is quite large, 
$(|\mu_{\rm eff}|^2 +m_{hu}^2)$ is just 
of order $M_Z^2$, particularly 
for large ${\rm tan}\beta$s.  
Hence, the EW breaking conditions 
\cite{book,HiggsPot},
\bea
&&|B\mu_{\rm eff}|^2 ~>~ 
(|\mu_{\rm eff}|^2+m_{h_u}^2)(|\mu_{\rm eff}|^2+m_{h_d}^2) ,
\label{EWbreak1} \\
&&2 |B\mu_{\rm eff}|~<~(|\mu_{\rm eff}|^2+m_{h_u}^2)
+(|\mu_{\rm eff}|^2+m_{h_d}^2)
\label{EWbreak2}
\eea 
are easily satisfied. 

In terms of $T_\zeta$, $\mu_{\rm eff}$ in \eq{effmu} 
is presented as 
\dis{ \label{muT}
\mu_{\rm eff}
\approx \frac{M T_\zeta}{\kappa}
\left(\lambda_2-\frac12\lambda_1T_\zeta\right)+\mu .
}
Particularly, if $|M\lambda_2/\kappa|^2$ 
is much larger (smaller) than $-m_{h_u}^2$, 
then $|T_\zeta|$
should dynamically be adjusted 
to a small (large) value, fulfilling \eq{extrm1}.
Although the bare $\mu$ is larger than $-m_{hu}^2$, 
it can still be true for $|M\lambda_2/\kappa|>|\mu|$ 
($>|\mu_{\rm eff}|$).
Note that in this case 
$|B\mu_{\rm eff}|$ 
would be much larger than $|B|\cdot |\mu_{\rm eff}|$.
%
Actually, the size of $T_\zeta$ depends on 
the given SUSY parameters such as $M$, $\mu$, 
$\kappa$, $\lambda_{1,2}$, etc. 
For $|T_\zeta|\ll 1$ ($|T_\zeta|\gg 1$), 
$H^2$ should decrease 
the coefficient of 
the quadratic (quartic) term 
of $\phi$ in the effective potential, 
i.e. $|\lambda_2H|^2+m_\phi^2\approx 0$ 
($|\lambda_1H|^2+m_X^2\approx 0$). 
%
For $|T_\zeta|\ll 1$ ($|T_\zeta|\gg 1$), thus,  
the Higgs VEV is simply given by 
$\sqrt{-m_\phi^2} ~/|\lambda_2|$ 
($\sqrt{-m_X^2} ~/|\lambda_1|$) in \eq{phiH}. 

Although $|m_{X,\phi}^2|$ remain light enough 
at the messenger scale where the gauge mediation 
effects come in the observable sector, however, 
they possibly become much heavier 
through their RG evolutions 
below the messenger scale.  
It is because the singlets, $X$ and $\phi$ 
are coupled to the MSSM Higgs fields in \eq{W}, 
and $|m_{h_{u,d}}^2|$ become quite heavy 
below the messenger scale 
by the gauge mediation effects of SUSY breaking.  
To keep the smallness of $|m_{X,\phi}^2|$, 
therefore, the coupling constants $\lambda_{1,2}$ 
need to be small enough and/or the messenger scale 
to be low enough. 
With a small enough $\lambda_2$ 
($\ll \lambda_1\lesssim 1$), e.g., 
we can get a sufficiently small 
$m_\phi^2$, and so we will attempt to explain 
the small Higgs VEV with $m_\phi^2$ in \eq{phiH}. 
Even if $m_X^2$ is quite sizable, its contribution 
to \eq{phiH} can still be suppressed 
by an extremely small $T_\zeta$ i.e. 
by a quite large value of $M$. 
Moreover, one can introduce other sizable couplings 
between $X$ and another heavy singlet sector 
such that $m_X^2$ is small at low energies,   
with leaving almost intact $m_\phi^2$ and 
the MSSM soft parameters.    
Thus, we will assume the first term 
is dominant in \eq{phiH}. 

\begin{figure}
\begin{center}
\includegraphics[width=0.9\linewidth]{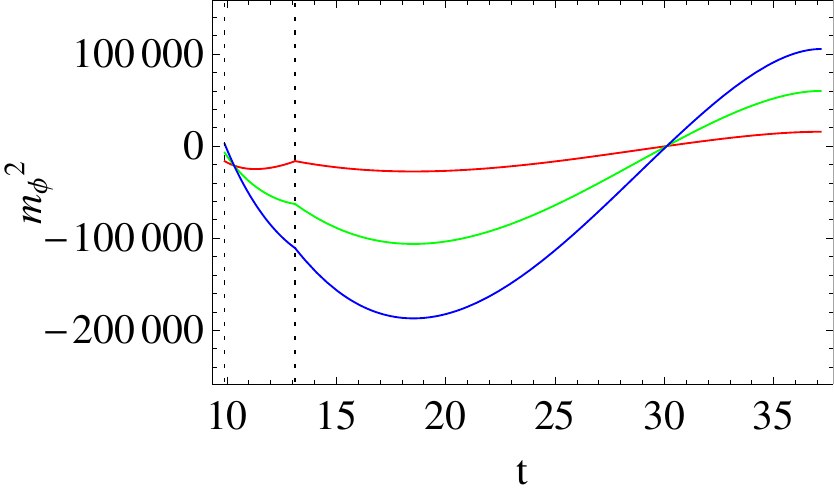}
\end{center}
\caption{RG evolutions of $m_{\phi}^2$ 
with $t$ [$\equiv {\rm log}(Q/\gev)]$
under the various trial universal squared soft mass
$m_0^2$s at the GUT scale. 
The right vertical dotted line indicates the messenger 
scale of $\Lambda_M=25 \tev$ ($10^6 \gev$)
in the Left (Right) figure, and 
the left vertical lines correspond to the 
stop mass scale of $\widetilde{m}_t=20 \tev$ 
in the both figures. 
In the left figures, the two dotted lines are 
almost overlapped.  
The other parameter choices in the both figures 
are the same as those of Case II.  
%
%
}
\label{fig:1}
\end{figure}

A relatively small value of $\lambda_2^2$ 
can make a ``focus point'' of $m_\phi^2$ emerge 
around the stop mass scale. 
The two figures in FIG.~\ref{fig:1} 
show the RG evolutions 
of $m_\phi^2$ under its various trial values 
at the GUT scale [$t={\rm log}(Q/{\rm GeV})
\approx 37$] with $\lambda_1^2=0.5$, $\lambda_2^2=8\times 10^{-3}$, 
and $\tb =40$.  
As seen in the both figures, 
two focus points of $m_\phi^2$ appear. 
The left vertical dotted lines indicate  
the $20 \tev$ stop mass scale, and 
the right vertical dotted lines correspond to 
the messenger scales of $25 \tev$ and $10^6 \gev$, 
respectively. 
In the first figure, the two dotted lines are almost 
overlapped because of the similarity of 
the stop mass and messenger scales. 
In the both cases, 
we set all the soft scalar masses being universal 
($\equiv m_0^2$) and all the ``A-term'' 
being the same as $m_0$ 
with the relatively heavier unified gaugino mass, 
$M_{1/2}=54 ~m_0$ at the GUT scale. 
It can be realized 
in ``no-scale'' SUGRA models \cite{book}.  
To keep the gauge coupling unification, here,  
we assumed the messenger fields compose one pair of  
$\{{\bf 5}, \overline{\bf 5}\}$ of SU(5),  
which are all decoupled below the messenger scale. 

%
%
\begin{figure}
\begin{center}
\includegraphics[width=0.9\linewidth]{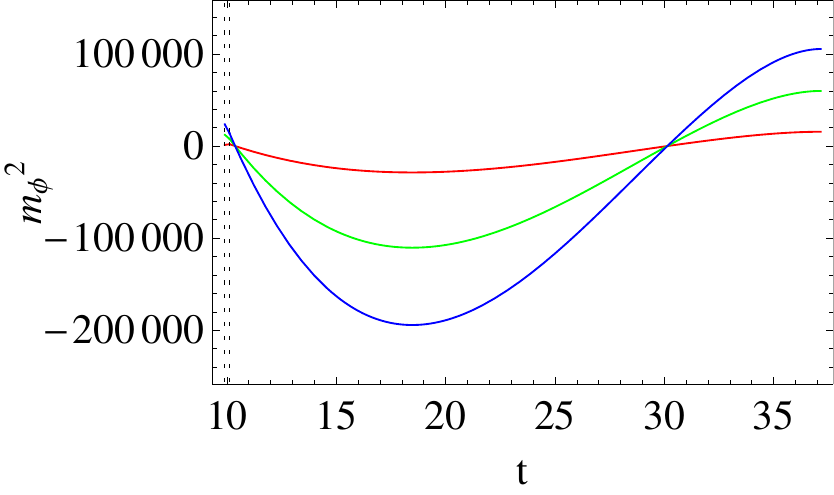}
\end{center}
\end{figure}
%
%
%
 %

Comparing the two figures, we see that  
the focus points of $m_\phi^2$ appear  
at the almost same energy scale,   
regardless of the messenger scales \cite{MyFP}. 
As a result, $m_\phi^2$s 
remain almost the same value at low energies, 
regardless of various trial values of $m_0^2$s 
at the GUT scale.    
In fact, larger (smaller) values of $\lambda_2^2$ 
and $M_{1/2}$ push the focus points 
to lower (higher) energy scales.  
Below the stop decoupling scale, 
the low energy value of $m_\phi^2$ can be estimated 
using the Coleman-Weinberg potential \cite{CQW}, 
where $m_{h_{u,d}}^2$ and $\mu_{\rm eff}$ 
dominantly affect $m_\phi^2$ via the $\lambda_2$ 
coupling. 
Even if they are not well-focused 
around the stop mass scale, however, 
the focusing of $m_\phi^2$
would not much be destroyed 
with a small enough $\lambda_2^2$. 

%
%
%

%
%
%
%
\begin{table}[!h]
\begin{center}
\begin{tabular}
{c|c||c|c}
\hline\hline
 {\bf Case I} & {\footnotesize $\tb=10$}  
 &  {\bf Case II}  & {\footnotesize $\tb=40$}  
\\
$\lambda_2^2=5\cdot 10^{-4}$ & $\widetilde{m}_t^2=(10 \tev)^2$ 
& $\lambda_2^2=8\cdot 10^{-3}$ & $\widetilde{m}_t^2=(20 \tev)^2$
\\
$\lambda_1^2=0.5$ & $\Lambda_{M}=15 \tev$ 
& $\lambda_1^2=0.5$ & $\Lambda_{M}=25 \tev$
\\ \hline
 {\footnotesize ${\bf \Delta_{m_0^2}}$}~ & {\footnotesize $~19.1$}  &  {\footnotesize ${\bf \Delta_{m_0^2}}$}~  & {\footnotesize $~79.6$}  
\\ 
 {\footnotesize ${\bf \Delta_{M_{1/2}}}$} &  {\footnotesize $~83.2$}  & {\footnotesize ${\bf \Delta_{M_{1/2}}}$}  & {\footnotesize $~28.6$} 
\\
 {\footnotesize ${\bf \Delta_{y_2^2}}$}~~ &  {\footnotesize $~59.7$}  &  {\footnotesize ${\bf \Delta_{y_2^2}}$}~~   & {\footnotesize $~56.5$} 
\\
 {\footnotesize ${\bf \Delta_{\rm GM}}$}~ & {\footnotesize $~37.1$}  & {\footnotesize ${\bf \Delta_{\rm GM}}$}~  & {\footnotesize $153.5$}  
\\
 {\footnotesize ${\bf \Delta_{\Lambda_{M}}}$} &  {\footnotesize $~6.0$}  &{\footnotesize ${\bf \Delta_{\Lambda_{M}}}$}  & {\footnotesize $~21.3$}
\\ \hline\hline
%
%
%
%
%
\end{tabular}
\end{center}\caption{Fine-tuning measures for 
the various input parameters defined in the text  
for the two different cases. 
}\label{tab:Delta}
\end{table}
%
%

In TABLE \ref{tab:Delta}, we list 
various fine-tuning measures 
$\Delta_X$ 
($\equiv |\delta {\rm log}M_Z^2/\delta {\rm log}X|$ 
\cite{FTmeasure}), where $X$ denotes various input 
parameters
for the two cases, in which the stop squared masses 
($\widetilde{m}_t^2\equiv\sqrt{m_{q_3}^2m_{u_3^c}^2}$)
are given by $(10 \tev)^2$ and $(20 \tev)^2$,  
where $m_{q_3}^2$ ($m_{u_3^c}^2$) denotes 
the squared mass of the SU(2) doublet (singlet) stop, 
and the messenger scales ($\Lambda_{M}$) 
are $15 \tev$ and $25 \tev$, respectively. 
We set $M_{1/2}=125~m_0$ and $54~m_0$, 
respectively, at the GUT scale.  
For dominant gauge mediation effects, here, 
we take large values of 
$F_{g}/16\pi^2\Lambda_{M}$ 
($\equiv {\rm GM}$), $5.1 \tev$ and $10.5 \tev$ 
in Case I and II, respectively. 
Since still $F_{g}\ll F_{\rm grav}$, however, 
the gravitino mass, $m_{3/2}$ is determined dominantly 
by the SUSY breaking source of the gravity mediation, 
$m_{3/2}\approx F_{\rm grav}/\sqrt{3}M_P\sim m_0
=30.4 \gev$ and $124.7 \gev$, respectively.
As expected, most of the fine-tuning measures are 
well-below 100, although the stop mass is  
around $10$ or $20 \tev$. 

The first and second generations of 
the colored superpartners  
must be much heavier than the stop. 
Moreover, the off-diagonal components of 
the squared mass matrices for the scalar partners 
are relatively small  
because they are generated only through 
the gravity mediations of SUSY breaking.    
Accordingly, unwanted FCNC processes are adequately 
suppressed in this setup.    

The three gaugino masses in Case I and II are 
quite heavy:
$({\cal M}_G, {\cal M}_W, {\cal M}_B)\approx 
(11.8 \tev, 4.8 \tev, 2.7\tev)$ and  
$(22.2 \tev, 9.1 \tev, 5.2 \tev)$, respectively, 
at low energy. 
Also the Higgsino mass, 
$\mu_{\rm eff}$ is basically  heavy, 
of order $-m_{h_u}^2$ in this model.
In Case I and II, it is $2.5$ and $2.3 \tev$, 
respectively. 
The lightest mass eigenstate of the singlet fermions  
comes mainly from  
the fermionic components of $\phi$ and $X$. 
Its mass turns out to be about 
$\kappa\langle Y\rangle/(1+T_\zeta^2)\approx 
-\frac{1}{2} T_\zeta^{2}(a^*-b^*)$, 
of order sub-GeV or lighter.    
It can play the role of dark matter \cite{on-going}. 



Now let us discuss the physical masses of 
the singlet scalars and 
their mixing angles with the SM Higgs boson. 
In this scenario, 
a light singlet scalar is essential 
for compensating the large contribution of 
$-m_{h_u}^2$ to \eq{extrm1}.
Since this mechanism works 
through the $\mu_{\rm eff}$ couplings in \eq{V}, 
a large mixing between the light scalar and 
the SM Higgs boson 
might be expected in this class of models.  
Such a large mixing would induce 
sizable invisible decay of the Higgs boson.
An important reason to introduce 
the several scalar fields is for avoiding it.  

Neglecting the $a$ and $b$ parameters 
for simple analysis, 
the squared mass matrix for the scalar fields 
in this model ($\equiv {\cal M}^2_S$)
takes the following form: 
\begin{eqnarray} 
\left(
\begin{array}{l}
\quad m_H^2 \qquad\qquad~~ \lambda_2 H \mu_{\rm eff} 
\qquad\qquad\qquad 
\lambda_1 H \mu_{\rm eff} 
\\
\lambda_2H \mu_{\rm eff} ~~
m_\phi^2+|\lambda_2H|^2+|\kappa\phi|^2  ~~ \lambda_1\lambda_2|H|^2 +\kappa\phi M 
\\
\lambda_1H \mu_{\rm eff} \quad~ \lambda_1\lambda_2|H|^2 +\kappa\phi M  
\quad m_X^2+|\lambda_1H|^2+|M|^2 
\end{array}
\right)
\nonumber 
\end{eqnarray}
in the basis of $\{H, \phi, X\}$.
Here $m_H^2$ collectively denotes 
the mass parameters in the MSSM Higgs sector.
Since $\langle Y\rangle$ is relatively small while 
$Y$ is quite heavy, we ignored $Y$ here.
For the solutions of Eqs.~(\ref{XY}) and (\ref{phi}) 
to be stable,  
all the eigenvalues of the above mass matrix, 
$\{m_1^2,M_2^2,M_3^2\}$  
must be positive definite around our solution. 
Since the sign of $m_{\phi}^2$ is flipped to be 
negative at low energies, 
the origin $\langle\phi\rangle=\langle H\rangle=0$ 
is made unstable, 
whereas the solution obtained above 
with non-zero VEVs becomes a stable point. 
Including the radiative corrections 
as well as the extreme conditions, 
Eqs.~(\ref{extrm1}) and (\ref{extrm2}), 
the smallest eigenvalue $m_1^2$ could be identified 
with the observed Higgs mass 
for $|m_H^2|$, $|\lambda_{1,2}H|$, 
$|m_\phi^2|\ll |M|^2$, $|\kappa\phi|^2$.  
On the other hand, the largest eigenvalue
would approximately be $|M|^2$ or $|\kappa\phi|^2$, 
depending on the solution of $T_\zeta$. 

The above squared mass matrix can be diagonalized 
into ${\rm diag.}(m_1^2,M_2^2,M_3^2)$,  
using the $3\times 3$ orthogonal mixing matrix 
given by
\begin{eqnarray}
{\cal O}_3=
\left(
\begin{array}{l}
\qquad c_1c_2 \qquad~~ -s_1 \qquad~~ -c_1s_2
\\
c_2c_3s_1-s_2s_3 \quad c_1c_3 \quad -c_3s_1s_2-c_2s_3
\\
c_3s_2+c_2s_1s_3\quad c_1s_3 \qquad c_2c_3-s_1s_2s_3
\end{array}
\right) ~~,
\end{eqnarray}
where $c_{1,2,3}$ and $s_{1,2,3}$ mean 
${\rm cos}\theta_{1,2,3}$ and 
${\rm sin}\theta_{1,2,3}$, respectively. 
Since the mixing angles between the MSSM Higgs sector 
and other neutral scalars 
should phenomenologically be 
suppressed \cite{PDG}, we need to show 
$|s_{1,2}|\equiv|\epsilon_{1,2}|\lesssim 0.1$, 
while $|c_{1,2}|\approx 1$.
%
%
%
%
${\cal O}_3^T\cdot{\rm diag.}(m_1^2, M_2^2, M_3^2)
\cdot{\cal O}_3$ ($={\cal M}^2_S$)
yields a symmetric matrix 
${\cal M}^2_{ij}$ ($={\cal M}^2_{ji}$) 
with the following elements: 
\bea
&&{\cal M}^2_{11}\approx 
M_3^2~\varepsilon_2^2+M_2^2~\varepsilon_1^2 +m_1^2 ,
\nonumber \\
&&{\cal M}^2_{12}\approx 
M_3^2 ~\varepsilon_2 ~{\rm sin}\theta 
+M_2^2 ~\varepsilon_1 ~{\rm cos}\theta 
-m_1^2 ~\epsilon_1 , 
\nonumber \\ \label{M2_ij}
&&{\cal M}^2_{13}\approx 
M_3^2 ~\varepsilon_2 ~{\rm cos}\theta 
-M_2^2 ~\varepsilon_1 ~{\rm sin}\theta 
-m_1^2 ~\epsilon_2 ,
 \\
&&{\cal M}^2_{23}\approx \Delta M_{32}^2 
\left(1-\bar{\epsilon}^2\right) 
~{\rm sin}\theta ~{\rm cos}\theta ,
\nonumber \\
&&{\cal M}^2_{22}
\approx \Delta M_{32}^2 ~{\rm sin}^2\theta + M_2^2
-\left\{\Delta M_{32}^2 ~{\rm sin}^2\theta
+\Delta m_{21}^2\right\} \epsilon_1^2 ,
\nonumber \\
&&{\cal M}^2_{33}
\approx \Delta M_{32}^2 ~{\rm cos}^2\theta + M_2^2
-\left\{\Delta M_{32}^2 ~{\rm cos}^2\theta
+\Delta m_{21}^2\right\} \epsilon_2^2
\nonumber \\
&&\qquad\quad 
-\Delta M_{32}^2 ~{\rm sin}2\theta ~\epsilon_1\epsilon_2 , 
\nonumber 
\eea
where $\theta$ means the mixing angle $\theta_3$. 
%
%
For simpler expressions, here we introduced 
the new parameters defined as 
\bea
&&\varepsilon_{2,1}\equiv\epsilon_{2,1} ~{\rm cos}\theta
\pm\epsilon_{1,2} ~{\rm sin}\theta ,
\\
&&\bar{\epsilon}^2\equiv
\frac12\left(\epsilon_1^2+\epsilon_2^2\right)
+{\rm tan}\theta ~\epsilon_1\epsilon_2
+\frac{2\Delta m_{21}^2 ~\epsilon_1\epsilon_2}
{\Delta M_{32}^2 ~{\rm sin}2\theta} ,
\nonumber \\
&&\Delta M_{32}^2\equiv M_3^2-M_2^2 ~,\quad {\rm and}
\quad ~\Delta m_{21}^2\equiv M_2^2-m_1^2 .
\nonumber 
\eea
%
%
%
%
%
%
%
%
%
Since ${\cal M}^2_{ij}$ is identified 
with the squared mass matrix ${\cal M}^2_S$ 
obtained above,  
%
$\{M, \kappa\phi, m_H^2, m_\phi^2, m_X^2, 
\lambda_1H, \lambda_2H, \mu_{\rm eff}\}$ 
in ${\cal M}^2_S$ 
can be expressed in terms of 
the mass eigenvalues and the mixing angles, 
$\{m_1^2, M_2^2, M_3^2; 
\epsilon_1, \epsilon_2, \theta\}$ 
by comparing their matrix elements, 
and using the two equations,  
Eqs.~(\ref{muT}) and (\ref{phiH}), 
by which $\kappa\phi/M$ ($=T_\zeta$) and 
$|\lambda_2H|^2$ are related to $\mu_{\rm eff}$ 
and $-m_\phi^2$, respectively. 

The ${\cal M}_{11}^2$ in \eq{M2_ij} 
%
actually contains the SM and heavy Higgs fields. 
When $|\lambda_{1,2}H|^2{\rm sin}2\beta{\rm cos}2\beta$ 
is relatively smaller than 
other elements of ${\cal M}^2_{S}$, 
the mixing angle between the SM and heavy Higgs 
is suppressed \cite{Jeong}. 
Moreover, a relatively small $A_{1,2}H$
decouples the heavy Higgs from the singlet sectors. 
In this case, $m_H^2$ in ${\cal M}_S^2$
[or ${\cal M}_{11}^2$ in \eq{M2_ij}] 
can be regarded as the physical SM Higgs mass. 
Including the quartic contributions, then,  
the light Higgs boson mass would approximately be 
\dis{ \label{HiggsMass}
M_Z^2{\rm cos}^22\beta 
+\left|\frac{\kappa\phi^2|\lambda_1|^2}{|M|^2}
\right|^2|H|^2+\Delta m_H^2 ,
}   
where the second term corresponds to 
the tree level contribution of the singlets 
to the Higgs mass, 
and the third term indicates the radiative correction. 
Then, the eigenvalue $m_1^2$ in \eq{M2_ij} 
should reproduce the measured Higgs mass 
[$\approx (125 \gev)^2$] \cite{PDG}. 
As is well-known, however, 
the first term in \eq{HiggsMass}, 
the tree level mass is too small to explain it. 
As seen in \eq{M2_ij}, moreover, 
$M_3^2\varepsilon_2^2$ and $M_2^2\varepsilon_1^2$ 
make negative contributions 
to the observed Higgs mass. 
Although there are many mechanisms to raise 
the Higgs mass \cite{MyHiggsMass}, 
just for simplicity, in this letter we will 
restrict our discussion to the cases that 
they are comparable to
the second term of \eq{HiggsMass}, i.e. 
\dis{ \label{(11)constraint}
\left|\frac{\kappa\phi^2|\lambda_1|^2}{|M|^2}
\right|^2|H|^2\approx 
M_3^2\varepsilon_2^2 + M_2^2\varepsilon_1^2 .
}  
Once we get somehow the stop mass 
of $10$--$20 \tev$, thus,  
we will regard the measured Higgs boson mass 
as being explained 
by the radiative correction $\Delta m_H^2$ 
at three-loop level as in the MSSM \cite{3-loop}.


The identifications of $(1,2)$ and $(1,3)$ components 
of ${\cal M}^2_S$ and ${\cal M}^2_{ij}$ give  
\dis{ \label{off-Diag}
&\lambda_2H\mu_{\rm eff}\approx 
M_3^2~\varepsilon_2 ~{\rm sin}\theta
+M_2^2~\varepsilon_1 ~{\rm cos}\theta ,~~~{\rm and}
\\
&\lambda_1H\mu_{\rm eff}\approx 
M_3^2~\varepsilon_2 ~{\rm cos}\theta 
-M_2^2~\varepsilon_1 ~{\rm sin}\theta ,
}
respectively. 
%
%
Thus, a quite large $\mu_{\rm eff}$ 
determined from \eq{extrm1} would result in 
a quite large mass eigenvalue $M_2^2$ 
or $M_3^2$.
For $|\mu_{\rm eff}|= 1.5$--$2.5 \tev$ 
($\sim -m_{h_u}^2$) 
and $\lambda_1\approx 0.7$, thus, 
$|\lambda_1 H\mu_{\rm eff}|$ is about 
$(430 \gev)^2$--$(550 \gev)^2$. 
In fact, $-m_{h_u}^2$ falls in such a range, 
when the stop mass is about $10$--$20 \tev$ 
at the messenger scales of $15$--$25 \tev$. 
Assuming 
$|M_2^2\varepsilon_1|\gg |M_3^2\varepsilon_2|$ and 
$|{\rm sin}\theta|\sim 1\gg |{\rm cos}\theta|$, 
$M_2=4.3$--$5.5 \tev$ with 
$\varepsilon_1\sim 10^{-2}$
can meet \eq{off-Diag}. 
Particularly, e.g. 
$(M_2, M_3)\sim (5 \tev, 500 \gev)$ 
and $(\varepsilon_1, \varepsilon_2)\sim (10^{-2}, 10^{-1})$ 
can fit also \eq{(11)constraint} 
as well as \eq{off-Diag}
for $|\lambda_2/\lambda_1|\ll 1$ and 
$|\kappa\phi^2/M^2|\sim 1$.
%
%
The ratio of the above two equations yields 
\dis{ \label{e1/e2}
%
\frac{\varepsilon_1}{\varepsilon_2}\equiv 
\frac{\epsilon_1{\rm cos}\theta-\epsilon_2{\rm sin}\theta}
{\epsilon_2{\rm cos}\theta+\epsilon_1{\rm sin}\theta}
=\frac{M_3^2}{M_2^2}~
\frac{\frac{\lambda_2}{\lambda_1}-{\rm tan}\theta}
{1+\frac{\lambda_2}{\lambda_1}{\rm tan}\theta} , 
} 
which is also a useful expression.

Writing $\kappa\phi M$ and $\lambda_1\lambda_2|H|^2$ 
with the mixing angle as 
\dis{ \label{delta M^2}
\kappa\phi M = \pm M_\Sigma^2 
{\rm sin}\theta ~{\rm cos}\theta , 
~~
\lambda_1\lambda_2|H|^2
=\pm \delta M^2 {\rm sin}\theta ~{\rm cos}\theta , 
\nonumber 
}
the new mass parameters, 
$M_\Sigma^2$  and $\delta M^2$ 
are related to the mass eigenvalues and mixing angles 
as 
\dis{
M_\Sigma^2+\delta M^2
\approx \pm\Delta M_{32}^2(1-\bar{\epsilon}^2) ,
}
from the identification of the $(2,3)$ components 
of ${\cal M}^2_S$ and ${\cal M}^2_{23}$.
For $\kappa\phi M\gg\lambda_1\lambda_2|H|^2$, thus, 
$M_\Sigma^2$ is relatively much larger than $\delta M^2$, 
$\kappa\phi M/\lambda_1\lambda_2|H|^2
=M_\Sigma^2/\delta M^2\gg 1$. 


Let us parametrize  
$\kappa\phi/M$, and $\lambda_1/\lambda_2$ 
using the angle variables, $\zeta$ and $\xi$:  
\dis{ \label{theta}
\frac{\kappa\phi}{M}\equiv {\rm tan}\zeta ,
~~ 
\frac{\lambda_2}{\lambda_1}\equiv {\rm tan}\xi ,
}
where ${\rm tan}\zeta$ ($=T_\zeta$) is determined by  
Eqs.~(\ref{extrm1}) and $(\ref{muT})$. 
%
%
%
%
Then, $|\kappa\phi|^2$ and $|M|^2$ are expressed as 
$\pm M_\Sigma^2 {\rm sin}\theta {\rm cos}\theta\times 
{\rm tan}\zeta$ and 
$\pm M_\Sigma^2 {\rm sin}\theta {\rm cos}\theta\times 
{\rm cot}\zeta$, respectively, 
while $|\lambda_{2 (1)} H|^2$ is given by  
$\pm\delta M^2 {\rm sin}\theta {\rm cos}\theta\times {\rm tan}\xi ~({\rm cot}\xi)$.   
%
As a result, 
$m_\phi^2$ and $m_X^2$ are identified as  
\bea \label{mphi}
&&m_\phi^2\approx m_2^2-\Delta m_{21}^2\epsilon_1^2
+\Delta M_{32}^2\left(
\bar{\epsilon}^2-\epsilon_1^2\right){\rm sin}^2\theta
 \\
&&~~~ \pm M_\Sigma^2 {\rm sin}\theta ~{\rm cos}\theta
\left({\rm tan}\theta-{\rm tan}\zeta\right) 
\nonumber \\  \label{mX}
&&~~~ \pm\delta M^2 {\rm sin}\theta ~{\rm cos}\theta
\left({\rm tan}\theta-{\rm tan}\xi\right) ,
\nonumber \\
&&m_X^2\approx m_2^2-\Delta m_{21}^2\epsilon_2^2
\\
&&~~~ +\Delta M_{32}^2\left\{\left(
\bar{\epsilon}^2-\epsilon_2^2\right){\rm cos}^2\theta
-\epsilon_1\epsilon_2 ~{\rm sin}2\theta \right\}
\nonumber \\
&&~~~ 
\pm M_\Sigma^2 {\rm sin}\theta ~{\rm cos}\theta
\left({\rm cot}\theta-{\rm cot}\zeta\right)  ~~
\nonumber \\ 
&&~~~ \pm\delta M^2 {\rm sin}\theta ~{\rm cos}\theta
\left({\rm cot}\theta-{\rm cot}\xi\right) 
\nonumber 
\eea
from the $(2,2)$ and $(3,3)$ components of 
${\cal M}^2_S$ and ${\cal M}^2_{ij}$.

As discussed above, the Higgs VEVs, 
$\langle h_u\rangle$ and $\langle h_d\rangle$ vanish 
in the SUSY limit, where all the soft mass parameters 
disappear. 
Even in the SUSY limit, however, $\phi$ and $X$
can still develop large VEVs: 
The flat direction along $\kappa\phi^2/2+MX=0$
becomes alive in this limit as seen in \eq{V}. 
Then, the $2\times 2$ block-diagonal part 
in ${\cal M}^2_S$, 
\begin{eqnarray}
\left(
\begin{array}{l}
|\kappa\phi|^2\quad \kappa^*\phi^*M
\\
\kappa\phi M^* ~\quad |M|^2
\end{array}
\right) 
= |M|^2 \left(
\begin{array}{l}
{\rm tan}^2\zeta \quad {\rm tan}\zeta 
\\
~{\rm tan}\zeta ~~\quad 1
\end{array}
\right) 
\end{eqnarray}
is diagonalized to 
${\rm diag}.(M_2^2,M_3^2)\approx{\rm diag}.(0,|M|^2)$ 
or ${\rm diag}.(|M|^2, 0)$ for $|\zeta|\ll 1$
with the mixing angles 
$\theta=\zeta$ or 
$\theta=\frac{\pi}{2}+\zeta$, respectively 
(and $\epsilon_{1,2}=0$).  
We are more interested in the second case. 
Even when all the soft SUSY breaking terms 
in ${\cal M}^2_S$ turned on, however, 
the results would be perturbed just slightly, 
since the soft parameters and the Higgs VEVs are 
relatively small.

Around $\theta=\frac{\pi}{2}+\zeta$, i.e. 
when $\theta=\frac{\pi}{2}+\delta\theta$ 
($|\delta\theta|, |\zeta|\ll 1$), 
we have ${\rm sin}\theta\approx 1-(\delta\theta)^2/2$, 
${\rm cos}\theta\approx-\delta\theta$, 
$M_2^2\gg M_3^2$, and so 
Eqs.~(\ref{mphi}) and (\ref{mX}) are   
approximated as 
\begin{eqnarray}
&&m_{\phi}^2\approx M_3^2-M_2^2\left[
\frac{\epsilon_1^2+\epsilon_2^2}{2}
+\left(\epsilon_1\epsilon_2
+\zeta-\delta\theta\right)\delta\theta\right] ,
\\
&&m_X^2\approx 
M_2^2\left[1-
%
2\epsilon_1\epsilon_2\delta\theta
-\epsilon_2^2-(1-\eta)\frac{\delta\theta}{\zeta}
-{\cal O}(\delta\theta^2)\right] , \qquad 
\end{eqnarray}
where we set $\eta\equiv M_3^2/M_2^2$ and 
${\cal O}(\delta\theta^2)\equiv
(\delta\theta)^2[1
-(2\delta\theta/\zeta+\zeta/\delta\theta)/3]$. 
For $m_\phi^2=m_X^2=0$ and $\epsilon_{1,2}=0$, thus, 
$\delta\theta$ and $M_3^2$ are determined to   
$\delta\theta=\zeta$ and $M_3^2=0$ as expected.
On the other hand, 
for non-zero but small enough $m_\phi^2$, $m_X^2$,
and $\epsilon_{1,2}$, 
the values of 
$\delta\theta$ and $M_3^2$ become relaxed 
as follows: 
\dis{ \label{Diag}
\delta\theta\approx
\frac{\zeta\left(1-\epsilon_2^2\right)}
{1-\eta +2\epsilon_1\epsilon_2\zeta}  
~~ {\rm and} ~~ 
M_3^2\approx M_2^2  \left[
\frac{\epsilon_1^2+\epsilon_2^2}{2} 
+\epsilon_1\epsilon_2\zeta\right] .
\nonumber 
}
For a given $\zeta$ and $\lambda_2/\lambda_1$, hence, 
$\varepsilon_1/\varepsilon_2$ can be determined 
from \eq{e1/e2}. 
If $\delta\theta\approx\zeta\sim{\cal O}(10^{-1})$, 
$(\lambda_2/\lambda_1){\rm tan}\theta\leq{\cal O}(1)$,
and $\epsilon_2\sim {\cal O}(10^{-2})$   
as considered above, 
$\epsilon_1$ and $M_3^2/M_2^2$ should be 
of order $10^{-1}$ and $10^{-2}$, 
respectively. 
Although we take an extremely small value of $\zeta$, 
$\epsilon_1$ and $M_3^2/M_2^2$ still stays around 
those values  
for $\epsilon_2\sim {\cal O}(10^{-2})$ and 
$\lambda_2/\lambda_1\sim {\cal O}(10^{-1}$--$10^{-2})$, 
unless $\lambda_2/\lambda_1$ is much smaller 
as well.
Since $M_2$ is approximately given by $M$, thus, 
a large enough SUSY parameter $M$   
($|m_\phi^2|,~|m_X^2| \ll -m_{h_u}^2
< |M|^2 < \widetilde{m}_t^2$)  
suppresses $\epsilon_2$ [by \eq{off-Diag}] 
as well as $\zeta$ [by \eq{muT}]. 
On the other hand, 
a larger (smaller) value of $\mu_{\rm eff}$
makes $\epsilon_2$ larger (smaller). 
%
%


In conclusion, we have proposed a scenario where  
the EW symmetry is broken by 
a negative soft squared mass 
of an MSSM singlet scalar. 
We have employed both the gravity- and gauge-mediated 
SUSY breaking scenarios, assuming the latter effects 
dominate over the former ones.
As a result, the MSSM SUSY particles can be much 
heavier than the experimental bounds and 
the FCNC phenomena are adequately suppressed. 
On the other hand, the naturalness associated 
with the EW symmetry breaking can be maintained.  
By introducing several singlets, 
a flat direction is admitted in the SUSY limit, and 
a large mixing between the Higgs boson and the singlet 
sector can be avoided, $|\epsilon_{1,2}|\ll 1$.        
The large effect of $-m_{h_u}^2$ on the Higgs VEV 
in the MSSM can be compensated dynamically 
by the flat direction,  
while the small curvature of the flat direction 
is compensated by the Higgs boson.     
  

\vspace{0.3cm}

\acknowledgments

{\it Acknowledgments} 
B.K. thanks Kwang Sik Jeong and Doyoun Kim 
for useful discussions. 
B.K. is supported by 
the National Research Foundation of Korea (NRF) funded by the Ministry of Education, Grant No. 
2016R1D1A1B03931151.



\begin{thebibliography}{99}

\bibitem{book} 
For a review, for instance, see 
  M.~Drees, R.~Godbole and P.~Roy,
  ``Theory and phenomenology of sparticles: An account of four-dimensional N=1 supersymmetry in high energy physics,''  Hackensack, USA: World Scientific (2004) 555 p; 
%
  S.~P.~Martin,
  Adv.\ Ser.\ Direct.\ High Energy Phys.\  {\bf 21}, 1 (2010)
  [Adv.\ Ser.\ Direct.\ High Energy Phys.\  {\bf 18}, 1 (1998)]
  [hep-ph/9709356].
  References are therein. 

\bibitem{Witten} 
  E.~Witten,
  Nucl.\ Phys.\ B {\bf 188}, 513 (1981).
  doi:10.1016/0550-3213(81)90006-7

\bibitem{MSSMunif}
  S.~Dimopoulos, S.~Raby and F.~Wilczek,
  Phys.\ Rev.\  D {\bf 24} (1981) 1681;
%
  C.~Giunti, C.~W.~Kim and U.~W.~Lee,
  Mod.\ Phys.\ Lett.\  A {\bf 6} (1991) 1745;
%
  U.~Amaldi, W.~de Boer and H.~Furstenau,
  Phys.\ Lett.\  B {\bf 260} (1991) 447;
%
  P.~Langacker and M.~x.~Luo,
  Phys.\ Rev.\  D {\bf 44} (1991) 817;
%
  J.~R.~Ellis, D.~V.~Nanopoulos and J.~Walker,
  Phys.\ Lett.\  B {\bf 550} (2002) 99
  [arXiv:hep-ph/0205336].

\bibitem{HiggsPot}
 M.~S.~Carena and H.~E.~Haber,
  Prog.\ Part.\ Nucl.\ Phys.\  {\bf 50} 63 (2003)
  [hep-ph/0208209].
 See also 
 A.~Djouadi,
  Phys.\ Rept.\  {\bf 459} 1 (2008)
  [hep-ph/0503173].
%

\bibitem{PDG}
M. Tanabashi {\it et al.} [Particle Data Group], 
Phys.\ Rev.\ D {\bf 98}, 030001 (2018). 


 
\bibitem{3-loop}
See, for instance, 
  R.~V.~Harlander, J.~Klappert and A.~Voigt,
  Eur.\ Phys.\ J.\ C {\bf 77}, no. 12, 814 (2017)
  doi:10.1140/epjc/s10052-017-5368-6
  [arXiv:1708.05720 [hep-ph]].

\bibitem{Kim-Nilles}
 J.~E.~Kim and H.~P.~Nilles,
  Phys.\ Lett.\  {\bf 138B}, 150 (1984).
  doi:10.1016/0370-2693(84)91890-2

\bibitem{axionRVW}
  H.~Baer, K.~Y.~Choi, J.~E.~Kim and L.~Roszkowski,
  Phys.\ Rept.\  {\bf 555}, 1 (2015)
  [arXiv:1407.0017 [hep-ph]].

\bibitem{FP}
J.~L.~Feng, K.~T.~Matchev and T.~Moroi,
  Phys.\ Rev.\ Lett.\  {\bf 84}, 2322 (2000)  [hep-ph/9908309].   
%

\bibitem{MyFP}
B.~Kyae and C.~S.~Shin,
  Phys.\ Rev.\ D {\bf 90}, no. 3, 035023 (2014)
  doi:10.1103/PhysRevD.90.035023
  [arXiv:1403.6527 [hep-ph]];
  %
  B.~Kyae,
  Phys.\ Rev.\ D {\bf 92}, no. 1, 015027 (2015)
  doi:10.1103/PhysRevD.92.015027
  [arXiv:1502.02311 [hep-ph]];
  %
  D.~Kim and B.~Kyae,
  Phys.\ Rev.\ D {\bf 92}, no. 7, 075025 (2015)
  doi:10.1103/PhysRevD.92.075025
  [arXiv:1507.07611 [hep-ph]].

\bibitem{splitSUSY}
  N.~Arkani-Hamed and S.~Dimopoulos,
  JHEP {\bf 0506}, 073 (2005)
  doi:10.1088/1126-6708/2005/06/073
  [hep-th/0405159].

\bibitem{CQW}
  M.~S.~Carena, M.~Quiros and C.~E.~M.~Wagner,
  Nucl.\ Phys.\ B {\bf 461}, 407 (1996) [hep-ph/9508343].

\bibitem{FTmeasure} 
  J.~R.~Ellis, K.~Enqvist, D.~V.~Nanopoulos and F.~Zwirner,
  Mod.\ Phys.\ Lett.\ A {\bf 01}, 57 (1986);
   R.~Barbieri and G.~F.~Giudice,
  Nucl.\ Phys.\ B {\bf 306}, 63 (1988).      

\bibitem{on-going}
  K.~Y.~Choi and B.~Kyae, work in progress.

\bibitem{Jeong} 
  K.~S.~Jeong, Y.~Shoji and M.~Yamaguchi,
  JHEP {\bf 1411}, 148 (2014)
  doi:10.1007/JHEP11(2014)148
  [arXiv:1407.0955 [hep-ph]].

\bibitem{MyHiggsMass}
See, for instance,   
  B.~Kyae and J.~C.~Park,
  Phys.\ Rev.\ D {\bf 86}, 031701 (2012)
  [arXiv:1203.1656 [hep-ph]];
    B.~Kyae and J.~C.~Park,
  Phys.\ Rev.\ D {\bf 87}, 075021 (2013)
  [arXiv:1207.3126 [hep-ph]];
    B.~Kyae and C.~S.~Shin,
  Phys.\ Rev.\ D {\bf 88}, no. 1, 015011 (2013)
  [arXiv:1212.5067 [hep-ph]];
    B.~Kyae and C.~S.~Shin,
  JHEP {\bf 1306}, 102 (2013)
  [arXiv:1303.6703 [hep-ph]];
    B.~Kyae,
  Phys.\ Rev.\ D {\bf 89}, no. 7, 075016 (2014)
  [arXiv:1401.1878 [hep-ph]].


\end{thebibliography}
\end{document}